\documentstyle[12pt,twoside,fleqn,espcrc1,epsfig]{article}

\title{A systematic study of two particle correlations 
from NA49 at CERN SPS}

\author{R.Ganz\address{Max-Planck-Institut f\"ur Physik, 
                       F\"ohringer Ring 6,
                       D-80805 M\"unchen,
                       Germany} for the NA49 collaboration} 

\begin{document}
\maketitle
\begin{abstract}
A systematic study of two particle correlations
measured by the NA49 experiment is summarized. 
Radii from Bose Einstein interferometry have been determined 
separately in different parts of phase space, 
for different collision systems
and at different incident beam energies.
Moreover, first results of a new method 
of accessing space-time asymmetries in the emission of particles 
by means of non identical particle correlations are presented. 
\end{abstract}

\subsection*{Bose Einstein Interferometry}
The aim of  studying two particle correlations is the 
reconstruction of the freeze out conditions of final state particles. 
In relativistic heavy ion collisions it appears 
that this state has to be described by a set of rather complicated 
non-isotropic, non-static and particle-type dependent emission functions.
To constrain these one has to gather selective information over a 
wide region of phase space by applying different analysis methods 
and by a systematic change of the initial condition of the collision.
The NA49 experiment~\cite{NIM99} 
with its four large volume TPCs detecting 
over 60\% of all pions emitted 
has collected rather a comprehensive body of data.
It comprises centrality selected Pb$+$Pb collisions at 158 and at 40~AGeV, 
minimium bias  Pb$+$Pb events at~158 AGeV as well as 158~AGeV p$+$p and p$+$Pb 
interactions. Overall this represents an ideal basis for such a study.
Usually freeze out condition are investigated by means of intensity 
interferometry which exploits correlations 
at small momentum differences $Q$ of two particles 
arising from quantum interference in the case of the two 
particles being indistinguishable. 
The description of the correlation function $C_2(Q)$ by a gaussian parametrization 
in the three components of the momentum difference 
$Q_{\mathrm{side}}$, $Q_{\mathrm{out}}$, $Q_{\mathrm{long}}$
relates the gaussian width parameters to the transverse ($R_{\mathrm{side}}$) 
longitudinal ($R_{\mathrm{long}}$) and
temporal ($R^2_{\mathrm{diff}}=R^2_{\mathrm{out}}-R^2_{\mathrm{side}}$) 
extent of the source at freeze out.
In order to observe the mere quantum statistical interference of 
like sign charged pions the correlation function has to be corrected 
for correlations owing to final 
state interactions mainly due to the Coulomb repulsion. 
For our central Pb$+$Pb 158~AGeV data this correction 
is done -- as described in \cite{NA35Coul} -- 
by the measured correlation function of unlike sign charged pairs.
\begin{figure}[hbt]
\begin{center}
\begin{minipage}[b]{0.6\linewidth}
\mbox{\epsfig{file=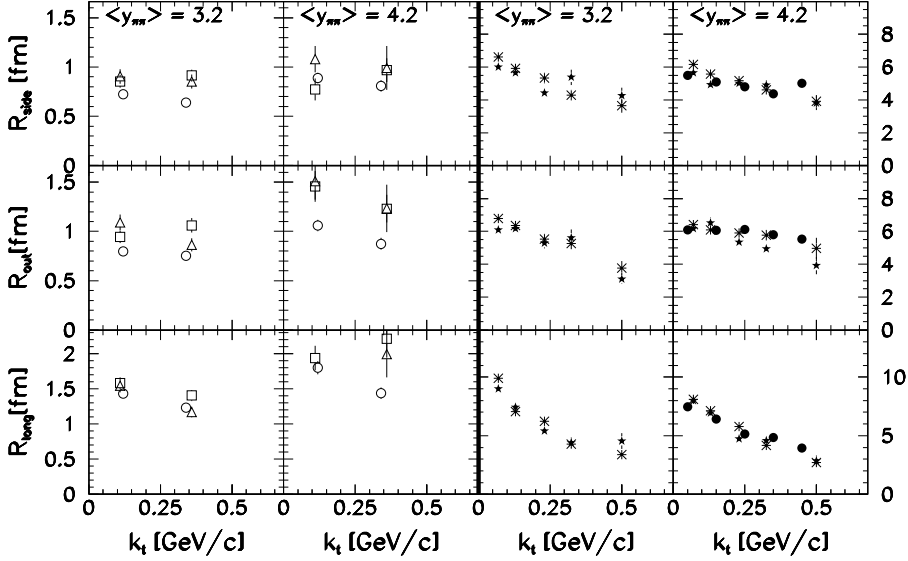,bbllx=80pt,bblly=300pt,bburx=348pt,bbury=407pt,width=0.65\linewidth}}
\end{minipage}
\hfill
\parbox[b]{.3\textwidth}{\sloppy
\caption{\small k$_{\mathrm{t}}$-dependence of radii. 
Two left columns: $\circ$ for p$+$p;  
$\Box$ for peripheral 
and $\triangle$ for central p$+$Pb collisions. 
Two right columns (extended vertical scale) 
are from separate analyses of central Pb$+$Pb collisions
for h$^-$h$^-$ ({\large $\bullet$}, {\large $\star$}) 
and for h$^+$h$^+$ ({\large$\ast$}) in the two TPCs
({\large $\bullet$} MTPC; {\large $\star$} and {\large$\ast$} VTPC).}
 \protect\label{fig:pppAAA}}
\end{center}
\end{figure}
In the case of the minimum bias data sample as well as in case of  
the central 40~AGeV data analysis the analytical ansatz 
of the Coulomb correlation, as suggested in \cite{Led98}, is
chosen instead. It was incorporated 
in the fitting procedure of the uncorrected correlation function, 
to take the size dependence of the Coulomb effect properly into account.
Despite the two different methods the results obtained from the central 
158 AGeV data sample coincide with those from the most central 
of five bins \cite{Glenn} in the minimum bias 158 AGeV data.
Due to the expected small source size 
the usual Gamov factor is applied in case of
p$+$p and p$+$Pb collisions to correct for Coulomb correlations, 
noting that compared to 
the uncorrected case no significant change of the radii is observed.
To eliminate the inefficiency of the NA49 TPCs 
for detecting pairs of two close tracks, 
those with distances of less than 2~cm in the Pb$+$Pb analysis 
or those with relative laboratoy momenta $Q_{\mathrm{x,y}}<20$~MeV/c and 
$Q_{\mathrm{z}}<100$~MeV/c (z beam axis)
in the p$+$p and p$+$Pb have been excluded from the analysis.
This restriction is imposed on pairs of particles from the same event (corr) 
as well as on those combined from different events (mix). The latter are used
as reference  sample (denominator) in the 
correlation function $C_2= N^{\mathrm{pair}}_{\mathrm{corr}}/N^{\mathrm{pair}}_{\mathrm{mix}}$.
All results presented here refer to the Longitudinal Co-Moving System (LCMS)
as the Lorentz frame, 
which is determined for each pair in such a way that the sum of the 
longitudinal momentum components of the pair vanishes.

Results of the analysis of the p$+$p collisions are shown in
the two left columns of figure~\ref{fig:pppAAA} together with results
from the analysis of central (CMD$\ge$7)\footnote{For a definition
of CMD see the contribution~\cite{Glenn} to this conference.}
and peripheral (CMD$<$7) p$+$Pb collisions. The phase space is 
subdivided into two regions of the average rapidity $y_{\pi\pi}$ of the two pions
at $y_{\pi\pi} = 3.9$
and into two of the average pion transverse momentum $k_{\mathrm{t}}$
at $k_{\mathrm{t}}= 0.25$~GeV$/$c.
In each column of figure~\ref{fig:pppAAA} $<y_{\pi\pi}>$ is the
average over all pairs contributing to the respective region. 
Similarily $<k_{\mathrm{t}}>$ are used as abscissae.
To increase the statistical significance the distributions from
pairs of negative hadrons are added to those of positive hadrons 
since the separate analyses agree within errorbars. 
It should be emphasized that for these collision systems
this is the first time such a phase space dependent three dimensional 
analysis has become feasible.
The radii from p$+$Pb tend to be slightly larger than those from p$+$p.
In all cases an ordering is observed 
$R_{\mathrm{long}}>$$R_{\mathrm{out}}>$$R_{\mathrm{side}}$, 
the latter being consistent with the observation of a finite duration 
of freeze-out. 
Moreover, the p$+$p radii at large values of k$_{\mathrm{t}}$ 
appear to be smaller than those at low k$_{\mathrm{t}}$. 
Whereas in central Pb$+$Pb reactions (two right columns of
figure~\ref{fig:pppAAA}) such a behaviour of the transverse radii
is attributed 
to a hydrodynamical expansion of the system \cite{NA49exp}
in p$+$p it seems more likely to be 
a characteristics of the decay pattern of resonances such as the $\rho$ 
which play an important role at this shorter length scale.
\begin{figure}[hbt]
\begin{center}
\begin{minipage}[b]{0.50\linewidth}
\mbox{\epsfig{file=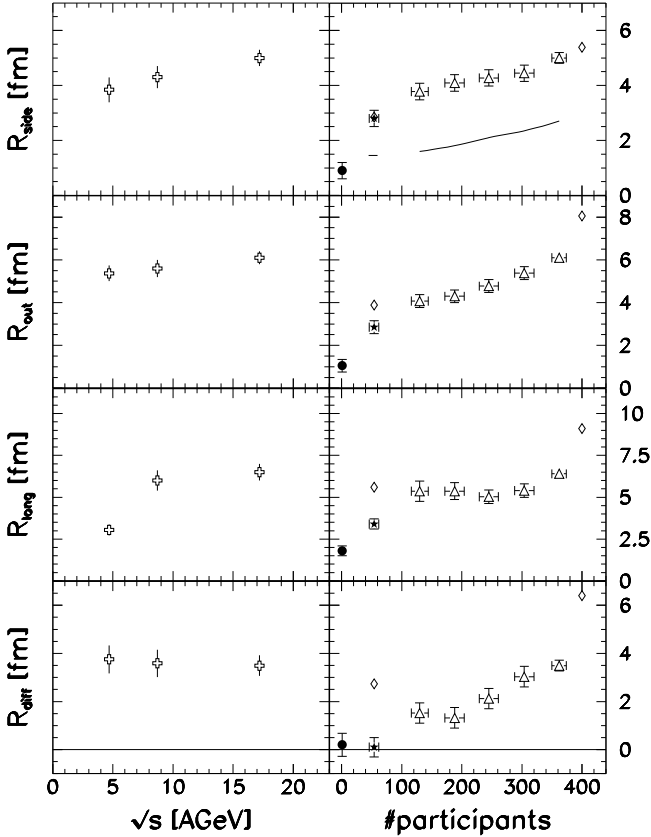,bbllx=49pt,bblly=190pt,bburx=540pt,bbury=430pt,width=\linewidth}}               
\end{minipage}
\hfill
\parbox[b]{.46\textwidth}{\sloppy
\caption{\small Left column: Dependence of radii on the center of mass energy. 
The 4.7 AGeV points are AGS Au$+$Au data from E866 $[7]$.
The 8.7 AGeV radii are our very preliminary 40~AGeV Pb$+$Pb results 
and the 17.2 AGeV points are for central 158 AGeV Pb$+$Pb.
Right column: Dependence of the radii on number of participants 
in the collision.
The $\bigtriangleup$ correspond to different centralities in 
158 AGeV Pb$+$Pb. 
Symbol {\large$\star$} is for central 200 AGeV S$+$S $[8]$
and {\large$\bullet$} is for the 158 AGeV p$+$p result. The $\diamond$ is  
the RQMD 1.08 result for central S$+$S 
and the RQMD 2.3 result for central Pb$+$Pb. 
The line$^2$ in $R_{\mathrm{side}}$ 
corresponds to the geometrical transverse size of the overlap region
of the two colliding nuclei.
($<y_{\mathrm{\pi\pi}}>\approx 4.2$ and $<k_{\mathrm{t}}>\approx0.12$~GeV$/$c .)}
 \protect\label{fig:centrality}}
\end{center}
\end{figure}
Complementary to the change of the collision system is the
variation in centrality of the collision. 
In the right column of figure~\ref{fig:centrality} the dependence on
the number $n_{\mathrm{part}}$ of nucleons participating in the collision
is shown for Pb$+$Pb collisions at different centralities
accompanied by our p$+$p result and by the NA35 result 
in the system S$+$S at 200~AGeV.
With increasing $n_{\mathrm{part}}$ 
the transverse radii $R_{\mathrm{out}}$ and $R_{\mathrm{side}}$
grow continously whereas $R_{\mathrm{long}}$ appears constant. 
$R^2_{\mathrm{diff}}$ 
is positive over the full range and $R_{\mathrm{diff}}$ 
grows linearly with $n_{\mathrm{part}}$. Similar to the findings 
of NA35 \cite{NA35SS} RQMD
gives a good agreement for $R_{\mathrm{side}}$ but 
overestimates $R_{\mathrm{out}}$ and $R_{\mathrm{long}}$. 
A comparison of $R_{\mathrm{side}}$ with an estimate 
of the geometrical transverse size of the overlap region 
in the collsions (figure~\ref{fig:centrality} 
line\footnote{The line in $R_{\mathrm{side}}$ 
corresponds to the average distance of points
in the overlap region of two spheres of radius $1.16 A^{1/3}$ 
offset by the impact paramter from the center of the the collision,
projected into a randomly oriented reaction plane.}) demonstrates 
that the interferometric radii reflect the freeze-out stage
which is preceeded by a strong expansion of the dense collision
zone produced shortly after the collision \cite{NA49exp}.
The left column of figure~\ref{fig:centrality} shows a first step
towards closing the gap between the highest AGS energy at 11.6 AGeV
and the SPS measurements at 158 AGeV by the 
(very preliminary) results from the 40~AGeV commissioning run 
of NA49 in 1998. Although the 40~AGeV and the 158~AGeV results are taken at 
$\frac{<y_{\mathrm{\pi\pi}}> - y_{\mathrm{cms}}}{y_{\mathrm{cms}}}
\approx 1.0$ whereas the AGS point \cite{E866} is measured at mid-rapidity, 
the comparison is justified by the observed slow
variation of the the radii with $y_{\mathrm{\pi\pi}}$ in NA49.
In the transition from AGS to the highest SPS energy 
$R_{\mathrm{long}}$ more than doubles 
whereas a small increase in the transverse radii 
$R_{\mathrm{out}}$ and $R_{\mathrm{side}}$ is observed  
and the temporal component $R_{\mathrm{diff}}$ even 
seems to be a constant.

\begin{figure}[hbt]
\begin{center}
\begin{minipage}[h]{0.45\linewidth}
\mbox{\epsfig{file=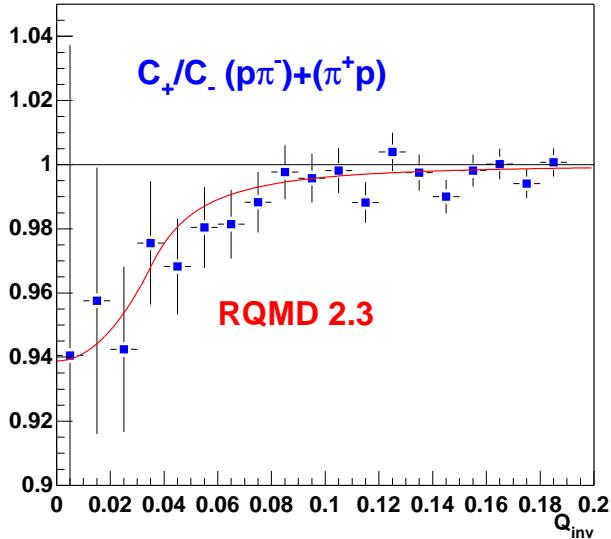,width=\linewidth}}               
\end{minipage}
\hfill
\parbox{.48\textwidth}{\sloppy
\caption{\small Ratio of correlation functions $C_+/C_-$ for central 158 AGeV  
Pb$+$Pb collsions (LCMS). To reduce statistical errors p$\pi^-$ and $\pi^+$p have been
combined. This is jusified by the consistency of $C_+/C_- \equiv 1$ in the case 
of $\pi^+\pi^-$ pair analysis. The solid line shows the assymetry
expected from the RQMD 2.3 event generator.}
 \protect\label{fig:nonid}}
\end{center}
\end{figure}

\subsection*{Correlations of Non Identical Particles}
A method for investigating space-time emission asymmetries 
has been suggested in \cite{Led96}, which utilizes  the effects of final 
state interaction on the correlations of two non identical particles 
at small relative velocities.
Contrary to the BE correlation method it gives access to the space-time 
asymmetries in the emission function of different particle species.
Such an analysis was carried out for the first time in the SPS energy regime and
is presented here for the case of proton-pion correlations.
The method involves the correlation functions $C_+(Q_{\mathrm{inv}})$ 
and $C_-(Q_{\mathrm{inv}})$ derived for the two
cases where $\cos\Psi>0$ and $\cos\Psi<0$, respectively, 
with $\Psi$ being defined as the angle between the pair velocity sum-vector 
and the velocity difference-vector in the rest frame of the pair
and $Q_{\mathrm{inv}}$ redefined as the momentum difference in the pair 
rest frame\footnote{In this frame $Q_{\mathrm{inv}}$ is equal to the relative velocity
times the reduced mass of the pair. For pairs of equal mass particles
the definition coincides with the usual definition of $Q_{\mathrm{inv}}$.}. 
A deviation of the ratio $\frac{C_+}{C_-}(Q_{\mathrm{inv}})$ from unity 
then allows -- besides a more quantitative evaluation --
for a distinction of emission patterns, for which 
on the one hand (I) low $p_t$ pions are emitted later 
or/and closer to the reaction axis and on the other hand (II) 
low $p_t$ pions are emitted earlier or/and further from the reaction axis 
than high $p_t$ protons.
Results for a small subset of  NA49 data of about 40.000 
central Pb$+$Pb events are shown in figure~\ref{fig:nonid}.
The data clearly exhibit asymmetries in the space-time emission 
favouring scenario I of the above two. 
Moreover, a good agreement is achieved in comparison to RQMD 2.3, in 
which the spatial component plays the dominant role in the occurance 
of the asymmetry observed for the model.

\end{document}